\documentclass[conference]{IEEEtran}
\IEEEoverridecommandlockouts
\usepackage{amsmath,amssymb,amsfonts}
\usepackage{algorithm}
\usepackage{algpseudocode}
\usepackage[left=1.62cm,right=1.62cm,top=1.9cm]{geometry}
\usepackage{array}
\usepackage[caption=false,font=normalsize,labelfont=sf,textfont=sf]{subfig}
\usepackage{stfloats}
\usepackage{url}
\usepackage{verbatim}
\usepackage{svg}
\usepackage{textcomp}
\usepackage{cleveref}
\usepackage{multirow}
\usepackage{graphicx}
\usepackage{xcolor}
\usepackage{booktabs}
\usepackage{enumitem} 
\usepackage{cite}
\usepackage{tabularx}
\usepackage{booktabs}
\usepackage{color}
\usepackage{soul}
\usepackage{array}
\usepackage{enumitem}
\usepackage{balance}
\usepackage{tikz}

\usepackage{listings}
\usepackage{xcolor}
\usepackage{adjustbox}

\definecolor{jsonbg}{rgb}{0.97, 0.97, 0.97}
\definecolor{jsonkey}{RGB}{0,0,150}
\definecolor{jsonstring}{RGB}{0,128,0}
\definecolor{jsonnumber}{RGB}{150,0,0}

\lstdefinelanguage{json}{
    basicstyle=\ttfamily\footnotesize,
    backgroundcolor=\color{jsonbg},
    numbers=none,
    showstringspaces=false,
    breaklines=true,
    frame=single,
    frameround=tttt,
    literate=
     *{0}{{{\color{jsonnumber}0}}}{1}
      {1}{{{\color{jsonnumber}1}}}{1}
      {2}{{{\color{jsonnumber}2}}}{1}
      {3}{{{\color{jsonnumber}3}}}{1}
      {4}{{{\color{jsonnumber}4}}}{1}
      {5}{{{\color{jsonnumber}5}}}{1}
      {6}{{{\color{jsonnumber}6}}}{1}
      {7}{{{\color{jsonnumber}7}}}{1}
      {8}{{{\color{jsonnumber}8}}}{1}
      {9}{{{\color{jsonnumber}9}}}{1}
      {:}{{{\color{black}:{}}}}{1}
      {,}{{{\color{black},}}}{1},
    keywordstyle=\color{jsonkey},
    stringstyle=\color{jsonstring}
}

\definecolor{delectricblue}{RGB}{100, 220, 255}

\def\BibTeX{{\rm B\kern-.05em{\sc i\kern-.025em b}\kern-.08em
    T\kern-.1667em\lower.7ex\hbox{E}\kern-.125emX}}
\usepackage[left=1.62cm,right=1.62cm,top=1.9cm]{geometry}

\begin{document}

\title
{A Secured Intent-Based Networking (sIBN) with Data-Driven Time-Aware Intrusion Detection}
\author{\IEEEauthorblockN{ Urslla Uchechi Izuazu, Mounir Bensalem, Admela Jukan}

\IEEEauthorblockA{\textit{Technical University of Braunschweig, Germany}} 

\{urslla-uchechi.izuazu, mounir.bensalem, a.jukan\}@tu-bs.de}
\maketitle

\begin{abstract}
While Intent-Based Networking (IBN) promises operational efficiency through autonomous and abstraction-driven network management, a critical unaddressed issue lies in IBN’s implicit trust in the integrity of intent ingested by the network. This inherent assumption of data reliability creates a blind spot exploitable by Man-in-the-Middle (MitM) attacks, where an adversary intercepts and alters intent before it is enacted, compelling the network to orchestrate malicious configurations. 
This study proposes a secured IBN (sIBN) system with data driven intrusion detection method designed to secure legitimate user intent from adversarial tampering. 
The proposed intent intrusion detection system uses a ML model applied for network behavioral anomaly detection to reveal temporal patterns of intent tampering. This is achieved by leveraging a set of original behavioral metrics and newly engineered time-aware features, with the model's  hyperparameters fine-tuned through the randomized search cross-validation (RSCV) technique.
Numerical results based on real-world data sets, show the effectiveness of sIBN, achieving the best performance across standard evaluation metrics, in both binary and multi classification tasks, while maintaining low error rates.
\end{abstract}


\vspace{-5mm}

\section{Introduction}
Technological advancements are driving networking toward distributed and autonomous systems with intent-based networking (IBN) emerging as a key enabler of this shift. IBN promises simplified operations by allowing network administrators to express high-level objectives rather than manually configuring devices \cite{leivadeas2022survey}. IBN adoption is rising rapidly, supported by standardization bodies like Open Networking Foundation (ONF) and 3rd Generation Partnership Project (3GPP),  and  is projected to grow from \$2.8 billion in 2024 to \$9.2 billion by 2030 \cite{ResearchMarketsIBN2025}. By abstracting low-level complexity, IBN is poised to form the operational core of next-generation technologies, like agentic AI, dynamically implementing complex network management workflows. Intent, a natural language expression of the desired workflows, is the core concept of IBN implementation, and can be simply seen as what the network administrator wants from the network management, without the need to handle the complexity of the underlying network infrastructure. 

Today's IBN systems operate on the  assumption that the intents presented for translation are authentic, authorized, and unaltered \cite{rebecchi2024revealing},\cite{kim2024security}. This creates a security gap, as the system becomes a force multiplier for an adversary (e.g., Man-in-the-Middle, MitM), who can subvert it at the intent level. A maliciously inserted intent will be automated and deployed at scale, leading to widespread network malfunctioning, service sabotage, or data breaches with minimal trace\cite{rebecchi2024revealing}. Furthermore, the proliferation of large language models (LLMs) for intent generation amplifies this threat. As LLM can easily produce either unintentionally erroneous, or deliberately malicious intents,  the existing IBN systems, though capable of validating syntactic or structural correctness \cite{jacobs2025establishing}, lack mechanisms for intent  integrity assessment before enactment in the network.

In this paper, we propose to secure an IBN system with a novel intrusion-detection system dedicated for intent integrity assessment. Moving beyond static inspection, our approach captures the behavioral footprints of compromised intents by analyzing their temporal impact on network operations. This approach, which we based on  real-world data set model, complements existing network security mechanisms, serving as a reliability checkpoint for validated intents prior to enactment within the network. The paper first conceptualizes a Man-in-the-Middle (MitM) attack  targeting the intent profiling phase, i.e., the initial phase where users express high-level objectives to the system. After that, it designs a security module based on randomized search cross-validation (RSCV) technique, used for effective detection of malicious intents. We implement a time-aware feature engineering strategy, enabling the proposed system to learn and detect behavioral fingerprints of manipulated intent. The real-world intent data set model, based on \emph{Business Intent and Network Slicing Correlation Dataset} (BINS) is used in a comprehensive performance evaluation across standard metrics for binary and multi-class tasks, showing the effectiveness of the proposed sIBN system.

The remainder of this paper is organized as follows: Section II reviews related work. Section III details the proposed system design. Section IV gives an overview of data and feature engineering implemented.  Sections V and VI presents the performance evaluation and conclusion, respectively.

\section{Related Work}

Related work in IBN has predominantly focused on improving functional capabilities, such as intent extraction/translation, with emerging work addressing policy conflict/resolution. Studies \cite{jacobs2018refining, el2023intent}, developed methods for capturing user intents via natural language interfaces and refining them using sequence-to-sequence models and operator feedback.  
Paper \cite{li2023policy} proposes a policy conflict detection method that uses translated network policies as the detection object, incorporates time constraint analysis, and handles both delayed and periodic intents. Similarly in \cite{zheng2022intent}, a conflict detector/policy resolution algorithm based on a two-dimensional analysis of intent endpoints and time spans was designed for an enterprise network.

Few efforts have been made in securing the IBN system.  Paper\cite{de2024novel}, presented a pioneering attempt towards intent security, whereby an intent detection system is proposed with a pre-defined security knowledge base to aid in the detection of known wrong intent expressions. Paper \cite{weintraub2024exploiting} demonstrated an exploitable temporal attack, i.e.,  \emph{phantom link attack}, where the order of flow rule installation is manipulated by delaying an intent at a selected switch. To counter this, spotlight was designed as a detection method to identify risky IBN updates.

All previous studies  assumed that data processed by the IBN is inherently trustworthy and reliable. Our paper is the first to address business-critical security challenge, where malicious actors could exploit this trust to introduce falsified intents. Authentication and authorization, while foundational, are not adequate to address this specific security challenge \cite{kim2024security}. They verify the identity of the entity submitting an intent, but this capability does not extend to scenarios where an intent is altered, or perhaps even more likely, generated by a permitted but compromised LLM agent. Similarly, rule-based systems offer protection against known, predefined threats, but are limited in their ability to detect novel attacks in form of deviations that characterize a malicious intent. 




\section{System Model}

\begin{figure*}[]
\centering 
\includegraphics[width=0.9\linewidth]{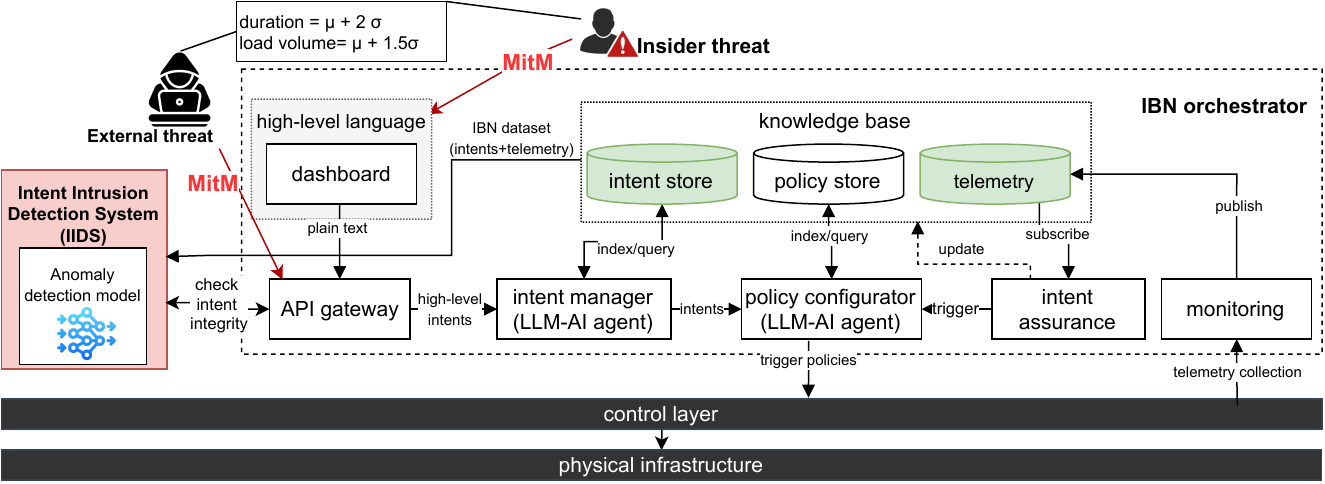}
\caption{Proposed Architecture for a secured Intent-based Networking (sIBN)} 
\label{IBN} 
\end{figure*}

The proposed sIBN system is conceptualized as a modular, pipelined architecture designed for secure intent network management, as shown in Fig.\ref{IBN}. We illustrate here an IBN orchestrator, a control layer and a physical infrastructure. The IBN orchestrator includes a dashboard, API gateway, intent manager, policy configurator, knowledge base,  monitoring, which are all standard IBN modules, and a newly proposed model called Intent Intrusion Detection System (IIDS), which is responsible for assessing the security of given intent, assuming the existence of malicious users.  

Without threats, the legitimate users can typically send business intents from the dashboard, which will be forwarded to the intent manager through an API gateway. The intent manager translates the business intent into structured  intent using predefined intent vocabularies. The intent manager can be implemented as an LLM-based AI agent which receives a high-level intent, a set of intents in forms of questions, containing natural language conditions, then respond with a structured intent. Afterwards, the policy  configurator  matches the user's requirements with existing policies stored in the policy store within the knowledge base. It checks for policy conflicts before mapping the intent into a lower-level, enforceable format. The policy configurator can also be implemented as an LLM-based AI agent that takes the structured intents, and list of tools configurable with API calls and NLP definitions stored in the knowledge base, then produce policies that can be enforced by the network controller. The intent assurance module is responsible for monitoring network telemetry, detects deviations from the intended state, and automatically triggers the policy configurator to enforce corrective actions, enabling proactive and intelligent network management. 

In the presence of threats, the adversary is modeled as a  MitM   aimed at  manipulating  legitimate intent $I$ to trigger misconfiguration. This threat can originate from two primary vectors: \textit{insider threats,} comprising malicious actors with authorized privileged access, and \textit{external threats}, with unauthorized access (e.g via compromised API), as illustrated in Fig.\ref{IBN}. To detect these threats, the newly proposed IIDS  analyzes the IBN datasets received from the knowledge base, and uses the trained models to check all the API calls to check for any tampered intent. If malicious intent is detected, it flags the request, thus only verified intents proceed downstream for processing. The IIDS operates on a structured data model that defines each intent through key parameters such as task loadvolume, duration, performance requirements etc. This system enables intent integrity checks, allowing the detection of deviations that indicate possible intent tampering.

\subsection{Threat Model}

The objective of an adversary \textit{(A)} is to alter or replace a stored intent with a tampered version \textit{I'} such that: 
\begin{equation}
    I' = f_A(I), \quad I' \neq I
    \tag{1}
\end{equation}

where \(f_A\) represents the adversary's manipulation function, which can alter one or more components of an intent (e.g., the network state description \(d\) or configuration metadata \(c\)). 
\vspace{2mm}

The ultimate goal of \textit{A} is to cause the IBN orchestrator(O) to implement a malicious network configuration:
\begin{equation}
    \text{O}(I') \rightarrow C_{\text{mal}}
      \tag{2}
\end{equation}
where \textit{C\textsubscript{mal}} denotes the malicious configuration resulting from orchestration of tampered intent \textit{I'}.

Equations~(1) and~(2) formally describe the adversary’s actions: how a legitimate intent \(I\) is transformed into a tampered intent \(I'\), and how this can lead the IBN orchestrator to produce a malicious configuration \(C_{\text{mal}}\), providing a formal basis for reasoning about intent tampering.

Attack  can occur at  various levels of the sIBN, however, this study focuses on the intent profiling phase, the initial phase, where users express their high-level network intent to the system. This phase represents a critical, yet under-explored attack surface where an adversary can manipulate submitted intents to achieve a high-impact compromise with minimal effort. MiTM attack is feasible at this stage due to the lack of an intent integrity checks at the API Gateway.  By anchoring security at the intent origin, we establish a foundational layer of trust essential for trustworthy IBN. This preemptive verification ensures that malicious data is prevented before it can corrupt the AI-driven automation pipeline, thereby directly contributing to the resilience of the entire system. Consequently, this approach is not merely an enhancement but a prerequisite for achieving reliable, large-scale network automation capable of operating dependably even under adversarial conditions.

\subsection{Intent Intrusion Detection System (IIDS)}

\begin{figure}[htbp!]
\centering
\includegraphics[width=1\linewidth]{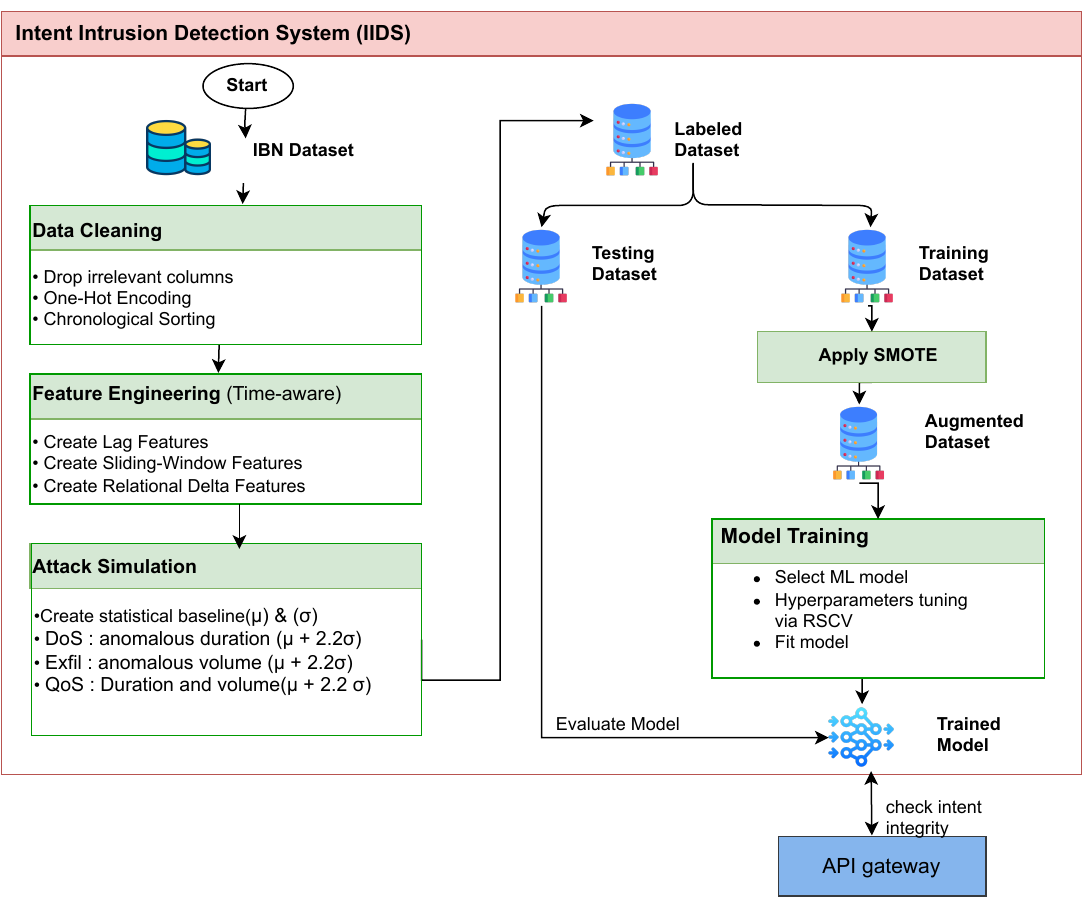}
\caption{Flowchart of the proposed Intent Intrusion Detection System (IIDS) 
}
   \label{IBN flow}
  \end{figure}
The proposed IIDS operates as an anomaly detection model within the IBN framework, focusing on the intent profiling stage, the point most vulnerable to tampering. Fig.\ref{IBN flow} illustrates the steps taken in designing the proposed IIDS. IBN data retrieved from  the knowledge base, form the input for the model training and real-time detection. This dataset undergoes a cleaning process and chronological sorting to maintain temporal consistency. Cleaned data is then subjected to feature engineering and attack simulation. The resulting labeled dataset is then split to training and testing, with the training set augmented using  synthetic minority oversampling (SMOTE) technique  to enable the model learn a more robust decision boundary. During training, suitable ML algorithms are selected, and hyperparameters optimized via RSCV. The trained model is then evaluated using the testing dataset. Finally, the model is integrated with the API gateway to monitor incoming intents, checking for intent integrity violations. We provide a detailed description of data and feature engineering in this workflow.

\section{Data and Feature Engineering}\label{sec:data}

In this section, we will focus on explaining the data preprocessing steps, which are data cleaning, feature engineering and attack simulation, as shown in Fig. \ref{IBN flow}. First, data preprocessing begins with systematic data cleaning, a critical step that addresses common data quality issues that can impede model performance. This involves handling missing values, duplicates, irrelevant fields, and attributes with high cardinality or zero variance. For instance, in dataset like CICIoT2023 \cite{neto2023ciciot2023}, attributes such as source IP  can cause excessive dimensionality if encoded, while constant fields add no analytical value.

Secondly, a step of feature engineering is needed  to create a more focused feature set that captures the behavioral fingerprints of network operations. In our work, we  engineered time-aware features to capture the fingerprints of normal network operations, allowing the detection of anomalies. In Fig. \ref{fig:json_event}, we illustrate a JSON example of a monitored event, and some general collected parameters.
\begin{figure}[!h]
\centering
\begin{adjustbox}{max width=1.0\linewidth}
\begin{lstlisting}[language=json]
{ "event_name": "Cloud Computing Services",
  "event_type": "Online Service",
  "event_arrival_time": "2023-07-23T02:00:00",
  "event_end_time": "2023-07-29T04:00:00",
  "event_duration": 526798000,
  "event_load_volume": 4595,
  "network_resources_consumed_per_device": "0.0715%",
  "network_functions_required": ["Wireless Access",
                    "Transmetropolitan Transmission",
                     "Core Network Processing"],
  "performance_requirements":{"bandwidth":">980Mbps",
                              "latency":"<49.4ms"}
}
\end{lstlisting}
\end{adjustbox}
\caption{Example of JSON representation for a  \textit{Cloud Computing Service} event.}
\label{fig:json_event}
\end{figure} 
We propose a set of \textbf{\textit{Lag Features}} that will provide the model with a  memory of immediate preceding network event. The features are: ``last\_event\_duration", ``last\_event\_volume", and ``time\_since\_last\_event",   to enrich each event with the context of the event that came just before it. 
Afterward, we created \textbf{ \textit{Sliding-Window Features}}, by computing statistical aggregates like ``avg\_duration\_in\_window" and ``sum\_volume\_in\_window" over a K-event window (eg. K=50), a size chosen to balance statistical stability against responsiveness to  short-term anomalies. This was done to provide the model with a macro-level view of normalcy, essential for detecting ``low-and-slow" attacks that manifest as a gradual drift from the established baseline. Then, to quantify an
event’s deviation from the recent service-specific norms, \textbf{\textit{Relational Delta Features:}} ``change\_from\_avg\_duration" and ``change\_from\_avg\_volume" were engineered. For instance, an event duration may not be extreme outlier in absolute terms, but its deviation from the recent moving average for that specific service type, becomes a clear anomaly indicator. 

Furthermore, to simulate a realistic attack scenarios, a small size of $M<<S$ 
 samples is selected, reflecting the fact that attacks occur infrequently in real-world systems, where $S$ is the size of dataset.
This process creates samples of the anomalous relationships between an intent's parameters and its behavioral consequences. For instance, an intent with  \textit{intent type }``communication service"  has a latency\_req of \textless 20ms and consistently result in a \textit{event duration} of approximately 100ms. The simulation creates a counter-instances where for example, an intent with "communication service", results in an anomalous \textit{event duration} of 150ms, teaching the model to recognize this statistically significant, yet less obvious, inconsistent relationship as the fingerprint of an intent tampering attack.
To achieve this, we first established a statistical baseline of normal network operation. This``normal behavior" is defined  by calculating the mean ($\mu$), which represents the average value of a feature, and the standard deviation ($\sigma$), which measures its expected range of variation. 





Three distinct attack 
scenarios were simulated using a data-driven approach, with attack parameters derived from the statistical properties of benign traffic.

\begin{itemize}
   
\item{DoS via Intent Exploitation (DoS):} Aims to exhaust network resources through prolonged occupancy. This was simulated by increasing \textit{event duration} to $\mu + 2.0\sigma$ while maintaining normal  event load volume. 
    
 \item{Data Exfiltration via Covert Channel (Exfil):} The goal of this attack is to move a large volume of data stealthily. This was modeled by elevating \textit{event load volume} to $\mu + 2.2\sigma$ while preserving normal event duration, creating high-throughput transfer patterns indicative of data exfiltration.
    
\item{Quality-of-service Degradation (QoS):} This simulates subtle service impairment by slightly increasing both \textit{event duration} and \textit{event load volume} to $\mu + 1.5\sigma$, reflecting the performance characteristics of degraded network paths with increased latency and retransmissions.

\end{itemize}

Subsequently, the resulting labeled IBN dataset is then split to training and testing, with the training set augmented via  SMOTE  to ensure the model learn a more robust decision boundary. During model training, suitable ML algorithms were selected, and hyperparameters fine-tuned via RSCV. This is because achieving best performance requires tuning critical parameters such as the learning rate ($\eta$), maximum depth, $\gamma$, and $\lambda$ etc. To efficiently explore this high-dimensional search space,  RSCV was employed, which samples parameter combinations from predefined distributions rather than exhaustively testing all possibilities. This significantly reduces computational cost compared to grid search \cite{subacsi2024comprehensive}, while maintaining the ability to identify near-optimal configurations.

Formally, the hyperparameter optimization problem is:
\begin{equation}
\theta^* = \underset{\theta \in \Theta}{\text{argmin}} \ \mathbb{E}[L(\theta)]
\tag{3}
\end{equation}

\noindent where $\Theta$ is the hyperparameter search space and $L(\theta)$ is the loss function for configuration $\theta$.

Table~\ref{tab:hyperparams} summarizes the optimal hyperparameters identified through RSCV for binary and multi-class tasks.
\begin{table}[htbp]
\caption{Optimal Hyperparameter for Proposed Anomaly Detection Model Based on RSCV}
\label{tab:hyperparams}
\centering
\resizebox{\columnwidth}{!}{%
\begin{tabular}{@{} >{\ttfamily}p{2.5cm} >{\raggedright}p{2.5cm} >{\centering}p{1cm} >{\centering\arraybackslash}p{1.5cm} @{}}
\toprule
\textbf{Hyperparameter} & \textbf{Description} & \textbf{Binary Value} & \textbf{Multi-Class Value} \\
\midrule
\texttt {n\_estimators} & Number of boosted trees & 300 & 500 \\
\midrule
\texttt {max\_depth} & Maximum tree depth & 3 & 4 \\
\midrule
\texttt {learning\_rate} ($\eta$) & Step size shrinkage & 0.01 & 0.05 \\
\midrule
\texttt {subsample} & Train data fraction per tree & 0.9 & 1.0 \\
\midrule
\texttt {colsample\_bytree} & Feature fraction per tree & 0.9 & 1.0 \\
\midrule
\texttt {reg\_alpha} ($\alpha$) & L1 regularization term & 0 & 0.3 \\
\midrule
\texttt {reg\_lambda} ($\lambda$) & L2 regularization term & 0.8 & 1.2 \\
\bottomrule
\end{tabular}
}
\end{table}

\section{Performance Evaluation}

The experiment was performed using python 3.9.7, with TensorFlow 2.9.1. Data preprocessing and analytical operations were supported by the pandas and numPy libraries.

In the rest of this section, we describe the used dataset, define the performance metrics and then present the results discussion.

\subsection{Data Preparation }

We utilized the BINS dataset \cite{li2025business}, a publicly available dataset for IBN research, aggregated from multiple sources to provide a rich source for data-driven network research. This study leveraged ``data source 1" from the BINS collection, consisting of 10,001 records of real-world operational data from Sichuan Telecom, China. This dataset was chosen for its authenticity as it covers a range of network demands and key performance metrics across business scenarios.  The structured records are direct representations of high-level intents. For example, a record with business name ``web browsing'', intent type ``network service'', start time ``20201230", end time ``20201231", and a performance requirement of ``latency $\leq$ 200 ms'', corresponds to the business intent: `Ensure that the latency for web browsing does not exceed 200 ms from december 30, 2020, to december 31, 2020' etc,. 
We simulate the three attacks as described in Section \ref{sec:data} on the dataset in \cite{li2025business}, by updating the values of \texttt{event duration} for DoS,  \texttt{event load volume} for Exfil, and both parameters for QoS Degradation. 

Table \ref{tab:original_features} outlines the original  features of the BINS, prior to feature engineering.  
Simulated attack data is added as described in Section \ref{sec:data}, where $M=30$. 
The resulting labeled dataset comprising 10,000 samples (9,970 benign and 30 attack instances), is then split into  train set (80\%) and test set (20\%) with  SMOTE technique applied to the train dataset, while the test dataset maintained its original imbalanced sample distribution of 2000 instances to ensure a realistic and unbiased evaluation of the model's true performance.
\begin{table}[h]
\centering
\caption{Original Features in the BINS Dataset}
\label{tab:original_features}
\resizebox{\columnwidth}{!}{%
\begin{tabularx}{1\columnwidth}{@{}lX@{}}
\toprule
\textbf{Feature Name} & \textbf{Description } \\ \midrule
\texttt{business name} & Name of the service. (e.g., VoIP, cloud computing, etc.).\\
\midrule
\texttt{intent type} & The type of service. (e.g.,online service, network service, etc.). \\
\midrule
\texttt{event arrival time} & event start time. \\
\midrule
\texttt{event end time} & event end time. \\
\midrule
\texttt{event duration} & The total elapsed time of the event. \\
\midrule
\texttt{event load volume} & The total data volume transferred during an event. \\
\midrule
\texttt{names of devices} &  The sequence of network devices in the event's path \\
\midrule
\texttt{network resources consumed } & The percentage of network resources utilized.\\
\midrule
\texttt{network functions} & Required network function details. \\
\midrule
\texttt{various performance reqs} & Performance requirements of network.(e.g.,"latency \textless 200 ms) \\ \bottomrule
\end{tabularx}
}
\end{table}
\subsection{Metrics}
The prediction performance of the chosen  ML-models was evaluated based on the following metrics:
\begin{itemize}
    \item \textit{Accuracy (Acc)}: Ratio of correctly classified instances to the total, where   TP is \text{true positives}, TN is \text{true negatives}, 
 FP is \text{false positives} and FN is \text{false negatives}     
    \item \textit{Precision (Prec)}: Correctness of positive predictions among all predicted positives.
   
    
    \item \textit{Recall (Rec)}: Proportion of true positives correctly found. 

    
  \item \textit{F1-score}: As the harmonic mean of precision and recall, F1-score better captures a model's behavior under changing precision and sensitivity.   
    
    \item \textit{Matthews Correlation Coefficient.(MCC)}: measures classification quality, effective for imbalanced datasets.


    \item \textit{Mean Squared Error (MSE)}: a loss function that measures how close predictions are to actual values.

    \item \textit{Time}: measures model's latency. An algorithm capable of executing a task within the shortest possible time meets stringent security requirements. 
\end{itemize}

\subsection{Results Discussions}
The results in Table \ref{tab:multiclass_summary} indicate that the proposed XGBoost demonstrates clear superiority, achieving the highest accuracy (99.98\%), recall (99.95\%), and precision (98.57\%) among all classifiers. While DT, LGBM and RF models also perform strongly, XGBoost had the highest F1-Score (98.50\%) and lowest MSE (0.007), confirming its balanced performance. Although LR has a faster inference time (18.2ms vs. 19.2ms), its lower F1-Score (83.60\%) reveals a trade-off, establishing the proposed XGBoost as the most effective in multi-class task.

\begin{table}[h]
\centering
\caption{Multi-Classification by Models}
\label{tab:multiclass_summary}
\resizebox{\columnwidth}{!}{%
\begin{tabular}{@{}llccccccc@{}}
\toprule
\textbf{Model} & \shortstack{\textbf{Time}\\(ms)} & \shortstack{\textbf{Acc.}\\(\%)} & \shortstack{\textbf{F1-Score}\\(\%)} & \shortstack{\textbf{Recall}\\(\%)} & \shortstack{\textbf{Prec.}\\(\%)} & \shortstack{\textbf{MCC}\\(\%)} & \textbf{MSE} \\ \midrule

 LR & 18.2 & 97.10 & 83.60 & 87.40 & 89.60 & 68.00 & 0.009 \\
 DT & 20.3 & 98.90 & 93.51 & 99.85 & 93.37 & 93.12 & 0.019 \\
 LGBM & 19.8 & 97.13 & 90.23 & 91.15 & 93.13 & 93.50 & 0.017 \\
 SVM & 21.3 & 89.80 & 62.50 & 62.50 & 62.50 & 64.01 & 0.012 \\
 RF & 30.1 & 99.75 & 98.10 & 98.13 & 98.19 & 91.00 & 0.090 \\
 \midrule
 \shortstack{\textbf{Proposed}\\\textbf{XGBoost}} & \textbf{19.2} & \textbf{99.88} & \textbf{96.50} & \textbf{99.95} & \textbf{98.57} & \textbf{94.81} & \textbf{0.007} \\
\bottomrule
\end{tabular}%
}
\end{table}

\begin{table}[h]
\centering
\caption{Binary Classification by Models}
\label{tab:Binary class_summary}
\resizebox{\columnwidth}{!}{%
\begin{tabular}{@{}llccccccc@{}}
\toprule
\textbf{Model} & \shortstack{\textbf{Time}\\(ms)} & \shortstack{\textbf{Acc.}\\(\%)} & \shortstack{\textbf{F1-Score}\\(\%)} & \shortstack{\textbf{Recall}\\(\%)} & \shortstack{\textbf{Prec.}\\(\%)} & \shortstack{\textbf{MCC}\\(\%)} & \textbf{MSE} \\ \midrule
LR & 6.01 & 91.70 & 85.72 & 89.02 & 89.36 & 71.31 & 0.009 \\
 DT & 18.9 & 99.07 & 92.01 & 94.08 & 94.00 & 91.67 & 0.003 \\

 LGBM  & 9.00 & 98.91 & 90.45 & 90.00 & 91.67 & 90.12 & 0.002 \\
  SVM & 25.0 & 89.65 & 65.13 & 62.91 & 66.36 & 65.02 & 0.009\\
  RF & 34.4 & 98.85 & 90.10 & 89.02 & 92.34 & 90.85 & 0.003 \\
 \midrule
 \shortstack{\textbf{Proposed}\\\textbf{XGBoost}} & \textbf{4.35} & \textbf{99.71} & \textbf{92.91} & \textbf{94.33} & \textbf{100.00} & \textbf{91.13} & \textbf{0.001} \\
\bottomrule
\end{tabular}%
}
\end{table}

\begin{figure}[htbp!]
\centering
\includegraphics[width=0.7\linewidth]{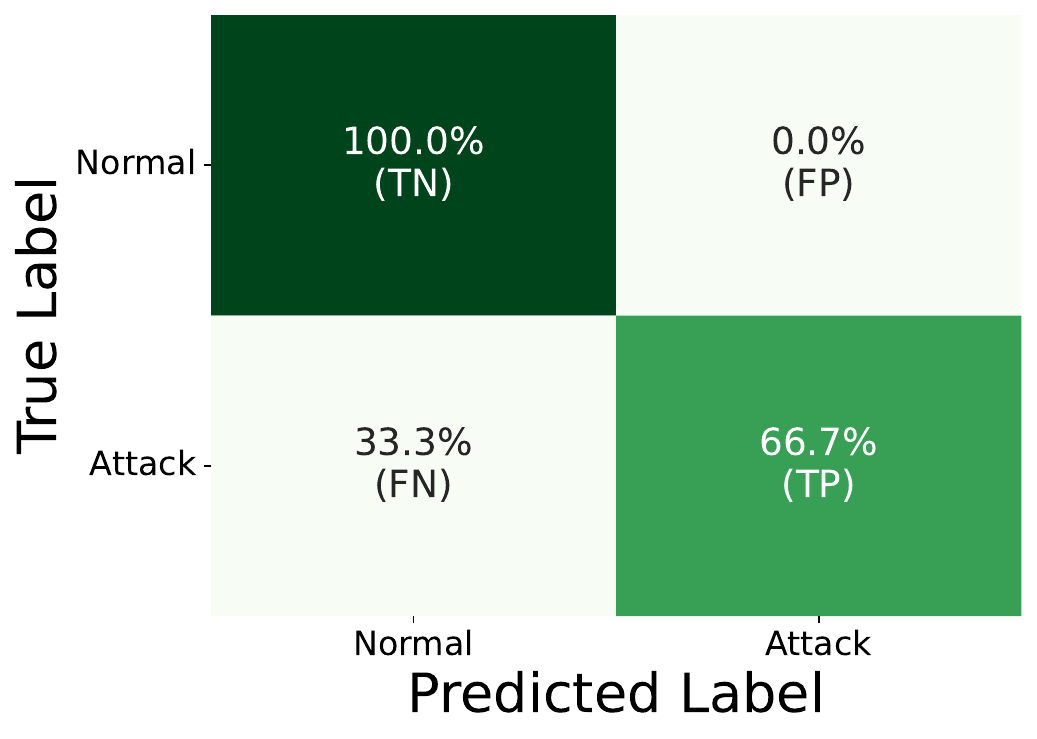}
\caption{Confusion Matrix based on for Binary and Multi-class Task}
   \label{confusion 1}
   \end{figure}

\begin{figure}[htbp!]
\centering
\includegraphics[width=0.8\linewidth]{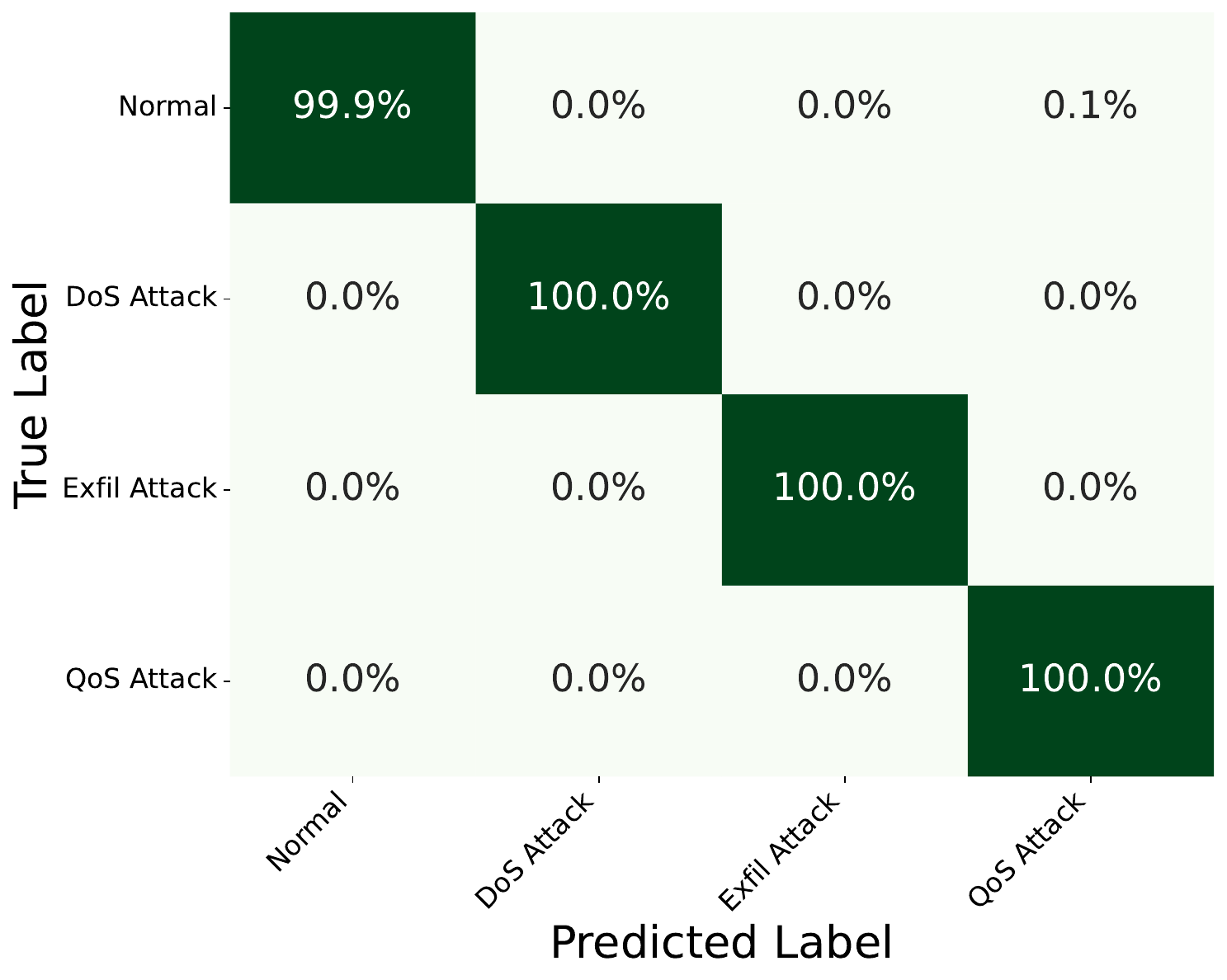}
\caption{Confusion Matrix based on  Multi-class Task}
   \label{confusion 2}
   \end{figure}

 Similarly, a binary classification evaluation was performed across all models, with results captured in Table \ref{tab:Binary class_summary}. Among the models, DT, LGBM, and RF achieved high accuracies above 98\% with good F1-scores, while SVM recorded the weakest performance. Notably, the proposed XGBoost gave the best overall performance, attaining an accuracy of 99.71\%,  100\% precision (indicating zero false alarms) , and the lest MSE of 0.001, while also achieving the shortest prediction time of 4.35ms. 
Furthermore, the confusion matrices in Fig.~\ref{confusion 1} and Fig.~\ref{confusion 2} show the percentage of TP, TN, FP, and FN for the multi-class and  binary classifications of the XGBoost model.

Based on the binary task, the proposed model show great performance, correctly classifying all 1994 normal instances (100.0\%)(TN), and 4 attack instances (66.7\%)(TP), while misclassifying 2 attack instance (33.3\%)(FN), and a 0.0\% FN. This highlights the model’s ability to distinguish between normal and malicious traffic with zero false positive.

Similarly, the multiclass confusion matrix demonstrates that the model classifies effectively  among the different scenarios, with accurate predictions reflected along the diagonal entries, despite a class imbalance that reflects real-world networks where normal traffic vastly outweighs attacks.

\section{Conclusion}
This study proposed to enhance IBN with  security features by introducing a security module  for intent integrity verification. The secure IBN (sIBN) system leverages behavioral and temporal features, with hyperparameters tuned via RSCV, and employs XGBoost as its core model. Simulation results confirmed superior performance over other models based on standard evaluation metrics. While the framework proved effective, the evaluation was conducted on a limited dataset, which may affect model generalizability. In the future, more sophisticated prompt attacks on IBN will be studied. We will also focus on expanding validation across diverse IBN environments, adapting the framework to a other networking use cases,  
and integrate explainable AI (XAI) techniques such as SHAP and LIME to improve transparency and trust in model prediction outcome. 


\balance
\bibliographystyle{./IEEEtran}
\bibliography{./IEEEabrv,./IEEE example,newref}
\end{document}